\documentstyle[12pt,aaspp4]{article}
\begin{document}

\title{Identification and photometry of globular clusters in M31 and M33
galaxies}
 
\author{B. J. Mochejska \& J. Kaluzny}
\affil{Warsaw University Observatory, Al. Ujazdowskie 4, 00--478 Warszawa,
Poland}
\affil{\tt e-mail: mochejsk@sirius.astrouw.edu.pl, jka@sirius.astrouw.
edu.pl} 

\author{M. Krockenberger, D. D. Sasselov \& K. Z.
Stanek\altaffilmark{2}}
\affil{Harvard-Smithsonian Center for Astrophysics, 60
Garden St., Cambridge, MA~02138}
\affil{\tt e-mail:krocken@cfa.harvard.edu, sasselov@cfa.harvard.edu,
kstanek@cfa.harvard.edu}
\altaffiltext{2}{On leave from N.~Copernicus Astronomical Center,
Bartycka 18, Warszawa 00--716, Poland}

\begin{abstract} 
We have used the data from the DIRECT project to search for new
globular cluster candidates in the M31 and M33 galaxies. We have found
67 new objects in M31 and 35 in M33 and confirmed 38 and 16 previously
discovered ones. A $VI$ and $BVI$ photometry has been obtained for all
the clusters in M31 and M33 respectively. Luminosity functions have
been constructed for the clusters in each galaxy and compared with
that of the Milky Way.  
\end{abstract}

\section{Introduction}

Many globular cluster searches have been conducted in the M31
galaxy. The first search, conducted by Hubble (1932), resulted in
the discovery of 140 globular clusters with $m_{pg}\leq 18$ mag. The
first major compilation of all the globular clusters known at the time
and their equatorial coordinates was compiled by Vete{\v s}nik (1962)
and contained about 300 objects. Since then that number has grown to
1028 objects appearing in at least one catalogue (Fusi Pecci et
al.~1993), of which only $\sim200$ are included in the three latest
major catalogues (Sargent et al.~1977, Crampton et al.~1985,
Battistini et al.~1980, 1987, 1993). Most of the previous searches had
been conducted on photographic plates using the technique of visual
identification of potential globular clusters. The catalogues are
fairly complete down to $V\sim18$ ($M_{V} \sim 6.5$), although the
degree of completeness is not uniform. According to Fusi Pecci (1993)
the catalogues are incomplete:

\begin{enumerate}

\item in the central part of the galaxy;

\item for bright, concentrated clusters, especially if superimposed on
a strong galaxy background;

\item for fainter objects.

\end{enumerate}

M33 had been searched for globular clusters rather sporadically. The
existence of globular clusters in that galaxy had been first noted by
Sandage in 1956 (Carnegie Institution Yearbook). The only recent
catalogue was compiled by Christian \& Schommer (1982), containing
around 200 objects.

Taking the above facts into consideration it seemed reasonable to
assume that new globular clusters could be found in both galaxies,
using the data collected by the DIRECT project (Kaluzny et al.~1998;
Stanek et al.~1998).  The frames obtained as part of that project
seemed suitable for the purpose of identification and photometry of
globular clusters for the following reasons:

\begin{enumerate}

\item large scale of the CCD frames (0.32 arcsec/pixel);

\item the limiting magnitude of the frames $V \sim 22$, in the light
of completeness of the M31 catalogues down to $V \sim 18$;

\item the frames covered central regions of the galaxies, where there
could still be undiscovered objects.

\end{enumerate}

\section{Data reduction}

All the observational data used here was taken from the DIRECT project
(Kaluzny et al.~1998; Stanek et al.~1998). Some additional data
generated by that project was also used, as will be later noted. The
frames for M31 were taken with the 1.3 McGraw-Hill telescope at
Michigan-Dartmouth-MIT (MDM) Observatory using the front-illuminated,
Loral $2048^2$ pixel CCD "Wilbur". At the f/7.5 station it had a pixel
scale of 0.32 arcsec/pixel and a field of view of about $11' \! \times
\! 11'$. Kitt Peak Johnson-Cousins $VI$ filters were used. The
observations for M33 were done with the 1.2 m telescope at the F.
L. Whipple Observatory (FLWO) using a thinned, back-side illuminated,
AR-coated Loral $2048^2$ pixel CCD ``AndyCam''. The pixel scale is the
same as in the case of M31. Standard Johnson-Cousins $BVI$ filters
were used.  The preliminary reduction of the frames was done as part
of the DIRECT project and will not be discussed here. Details on that
procedure can be found in Kaluzny et al.~1998. Bad columns were masked
out using the IMREPLACE routine of the IRAF\footnote{IRAF is
distributed by the National Optical Astronomy Observatories, which are
operated by the Association of Universities for Research in Astronomy,
Inc., under cooperative agreement with the NSF.} package.

For each of the fields in M31 ten frames were selected in $VI$ filters
with sub-arcsecond seeing. Exceptions were made for field C in the $V$
filter, where only seven frames were chosen, and for field D in the
$I$ filter, where only one frame was available.

The frame with the best seeing and lowest background was chosen as the
template for each of the fields. The choice of the template frames was
taken from the DIRECT project, as were the data used in creating lists
containing the positions of common stars on the frames. The
individual, non-template frames were transformed to the template
coordinate system with the GEOMAP and GEOTRAN routines of the IRAF
package. The frames were then averaged together using IMCOMBINE.

In the case of M33 single frames were used because of the substantial
amount of bad columns on the detector. Since each frame was offset by
a couple of pixels from the template, bad columns would cover a large
area on the combined image, thus hindering the identification of new
globular cluster candidates.

Photometry was extracted using the {\it Daophot/Allstar} package
(Stetson 1987). A point spread function (PSF) varying quadratically
with the position on the frame was used for the fields in M31. The PSF
was modeled with a Moffat function. Stars were identified using the
FIND subroutine and aperture photometry was done on them with the PHOT
subroutine. Around 100 bright isolated stars were chosen for the
construction of the PSF. The same stars were used as in the DIRECT
project, with the exception of the M31D field in $I$ filter.

For the fields in M33 the PSF was approximated by a Moffat function
linearly varying with position on the frame. Only about 50 stars were
used to construct the PSF and that proved to be sufficient.

The construction of the PSF consisted of two separate stages:

\begin{enumerate}

\item Using a modified version of the PSF-subroutine a preliminary PSF
was constructed in an iterative process. In each iteration the PSF
stars with profile errors greater than twice the average were removed
from the list.

\item The neighbors of the PSF stars were fitted using {\it Allstar}
and then subtracted with the SUBSTAR subroutine. An improved PSF was
constructed from the subtracted frame. This procedure was repeated
twice.

\end{enumerate}

The PSF obtained using the above method was then used by {\it Allstar}
in profile photometry. In case there were stars remaining on the
subtracted frame, FIND was ran again to identify them. PHOT was used
to determine the aperture photometry for the newly found stars. The
output file was combined with the one containing the stars found
previously and used as input to {\it Allstar}. In the case of fields
A, B in M33 it was found necessary to introduce another step, where
stars less than 3 pixels apart were removed and {\it Allstar} was ran
again on that file. The reason for this was that the second FIND had
been ran with a rather low threshold. Such a procedure was used to
make sure that the residuals of previously subtracted stars wouldn't
be identified and then fitted as separate objects.

\section{Photometric and astrometric calibration}

For all fields in M31 and fields A, B in M33 the photometric
calibration from the DIRECT project was used. The transformation to
the standard system was derived from observations of the Landolt
fields (Landolt 1992). More details on that subject can be found in
Kaluzny et al.~(1998).

In the field M33C the same calibration was used as for the field
B. This was possible because those fields overlapped by a substantial
amount and enough bright and isolated common stars could be found in
each filter.

\begin{figure}[htbp]
\vspace{8.5cm}
\includegraphics{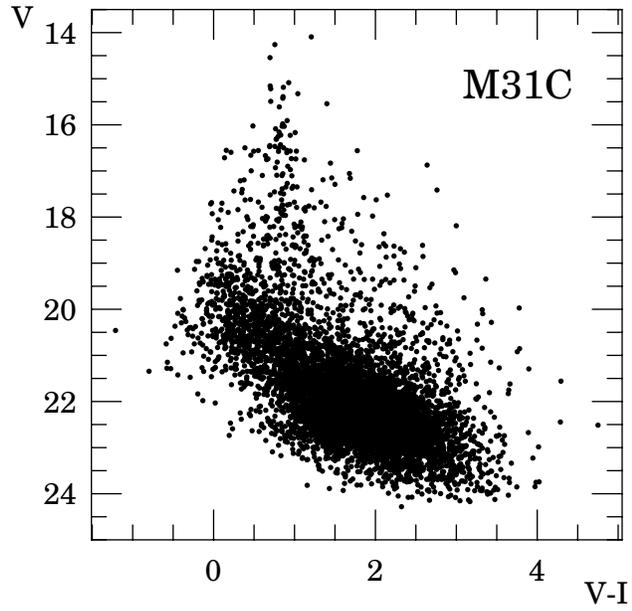}
\caption[]{The $V/V-I$ color-magnitude diagram for the stars in field C in
M31}  
\label{m31c}
\end{figure}

\begin{figure}[htbp]
\vspace{8cm}
\includegraphics{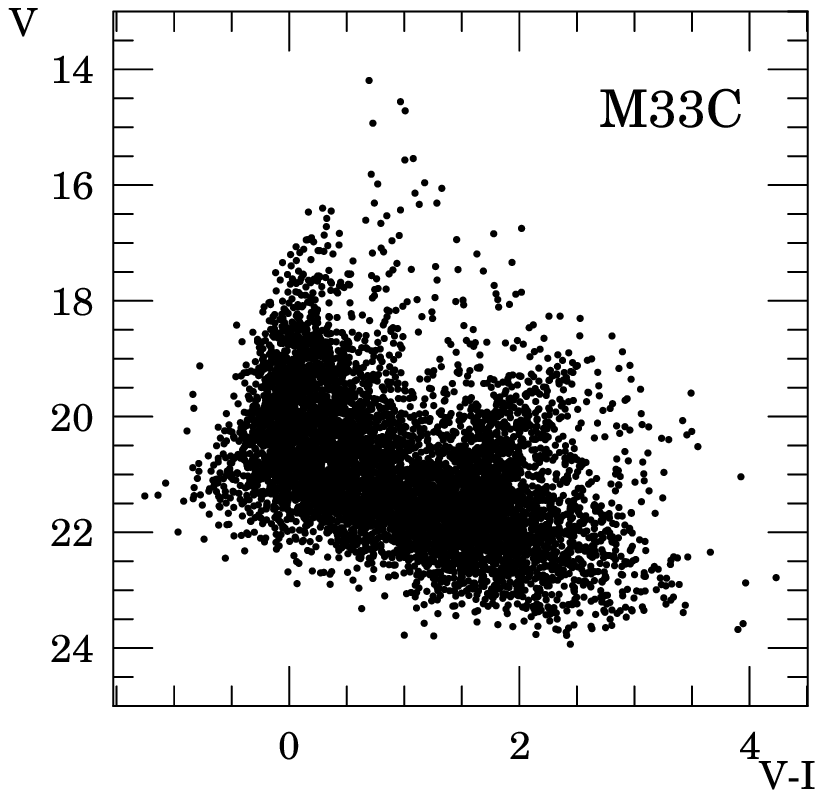}
\caption[]{The $V/V-I$ color-magnitude diagram for the stars in field C in
M33} 
\label{m33c}
\end{figure}

\begin{figure}[htbp]
\vspace{8cm}
\includegraphics{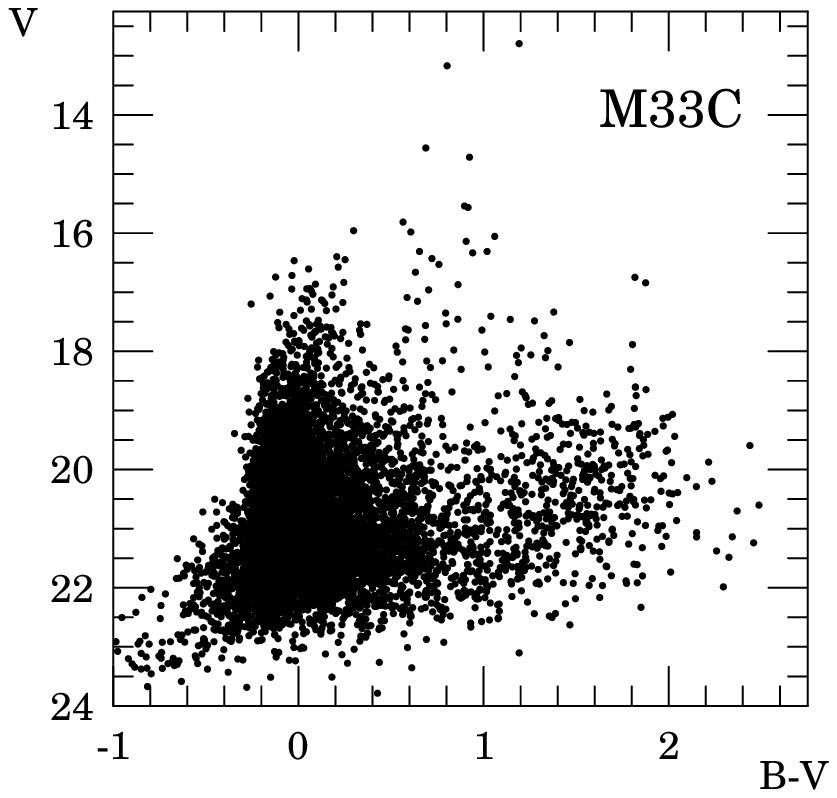}
\caption[]{The $V/B-V$ color-magnitude diagram for the stars in field C in
M33} 
\label{m33cBV}
\end{figure}

Sample color-magnitude diagrams are shown in Figs. \ref{m31c} -
\ref{m33cBV}.

The equatorial coordinates of globular clusters in M31 were derived
from the data generated by the DIRECT project. In the M33 galaxy the
right ascension and declination of the objects was found using
reference stars from the USNO A-1 astrometric catalogue. In both cases
the accuracy of the positions is better than 2 arcsec.

\section{Identification and photometry of GC candidates}

The following catalogues were used to identify the already known
globular clusters in M31: Battistini et al.~(1987); Crampton et
al.~(1985); Sargent et al.~(1977).  All objects from those catalogues
which were located within the studied fields had photometry done on
them, regardless whether they appeared to be globular clusters or not.

In the case of M33 the only catalogue used for this purpose was the
one compiled by Christian \& Schommer (1982). Magnitudes were measured
only for those objects which were identified independently in this
work.

The basic idea behind the identification of new globular cluster
candidates was the fact that the point spread function of a globular
cluster is different from that of a star. The PSF of a globular
cluster should have a larger FWHM than a stellar one. Taking that into
account it was expected that on the subtracted frames such objects
should exhibit toroidal residuals, thus enabling their identification.
Objects displaying this characteristic were selected and then their
nature was verified on the original frames. The following criteria
were used to discern globular clusters from other objects:

\begin{enumerate}

\item FWHM of the object had to be larger than that of a separate star;

\item the object had to be spherical.

\end{enumerate}

Objects which didn't meet those criteria in all filters were rejected.
Remaining objects were classified on a scale from A to D,

\begin{itemize}

\item[A] -- very high probability globular cluster candidates,
exhibiting spherical structure and a substantially larger FWHM than
for a nearby star.

\item[B] -- high probability globular cluster candidates. Objects in
this category usually exhibited small departures from spherical shape
not associated with the object itself, caused by nearby faint stars,
brightness fluctuations of the galaxy background, etc.

\item[C] -- objects that might be globular clusters. Such objects were
usually faint and roughly spherical. Due to their faintness their
radial profiles showed substantial background variations. Thus it was
hard to discern whether they were globular clusters or blended stars.

\item[D] -- objects that are probably not globular clusters. Objects in this
category met the above criteria rather poorly, usually because of low S/N
ratio. Objects from other catalogues which were found unlikely to be globular
clusters were also put into this category. 
\label{klasy}

\end{itemize}

It should be noted that the criteria used in the selection of globular
cluster candidates were not very keen in discerning globular clusters from
other non-stellar objects, especially if faint. HII regions, planetary
nebulae are some of the possible sources of false identifications.

This classification is similar to the one used in the Battistini
(1987) catalogue. One important difference is that for objects in
classes A and B Battistini put on an additional restriction on their
color.

After having created the list containing all globular cluster
candidates photometry was done on them. This procedure consisted of
several separate steps:

\begin{enumerate}

\item Selection of reference stars\\ 
Several (3-5) reference stars were selected for each frame in order to
convert the instrumental magnitudes of globular clusters to the
standard system. For this purpose bright stars with known standard
magnitudes were selected, located in regions with low background
brightness.

\item Subtraction of neighboring stars\\
Most of the globular clusters were rather faint, so it was necessary
to subtract nearby stars in order not to overestimate their
brightness.

\item Photometry of globular clusters\\
Since the point spread function of a globular cluster differs
substantially from the stellar PSF derived by Daophot and later used
by {\it Allstar}, profile photometry wasn't suitable for this
purpose. Aperture photometry was used instead. Magnitudes were
measured inside an aperture of 18 pixels using the PHOT
subroutine. PHOT had problems determining the magnitude when there
were bad pixels near a cluster, resulting from the subtraction of
nearby faint objects. In such cases those pixels were filled with the
value of the median computed in a box of $11 \! \times \! 11$
pixels. Bad columns were treated in a similar fashion. Magnitudes of
objects, which incorporated the use of this method are marked in the
tables by a colon.

\item Transformation to the standard system\\
The standard magnitudes were determined by the following formula:
\begin{equation}
M_{gc}=\frac{1}{N}\sum_{i=1}^{N} (m_{gc}-m_{ref}+M_{ref})
\end{equation}
where $m_{gc}$ and $m_{ref}$ are the instrumental magnitudes of a globular
cluster and a reference star respectively, and $M_{gc}$ and $M_{ref}$ are
their magnitudes referred to the standard system. N is the number of
reference stars. 

\end{enumerate}

Magnitudes obtained in the following manner were compared with results
from other catalogues. The degree of consistency was found to be
satisfactory. A more detailed comparison will be presented in section
\ref{cmpo}.

The photometric data for the globular cluster candidates, along with their 
equatorial coordinates and finding charts are available through {\tt
anonymous ftp} on {\tt cfa-ftp.harvard.edu}, in {\tt pub/kstanek/DIRECT}
directory.

\section{Conclusions}

\subsection{Accuracy of the photometry}

Our final result was the identification of new globular cluster
candidates in the M31 and M33 galaxies, the confirmation of the
previously discovered ones and the determination of their magnitudes,
using aperture photometry, in $VI$ and $BVI$ bands, respectively.

In crowded fields the aperture photometry is less accurate than the
profile photometry. As was mentioned earlier, the latter method of
determining the magnitudes could not be used because the globular
clusters could not be properly fitted with a stellar point spread
function.

In order to estimate the errors of the magnitudes found using profile
photometry a comparison was made between objects found in more than
one field. The magnitudes of already discovered clusters, determined
here, were compared with values taken from other catalogues.

\subsubsection{Comparison within the catalogue}

\begin{figure}[htbp]
\vspace{7.5cm}
%\hspace {-2cm}
\includegraphics{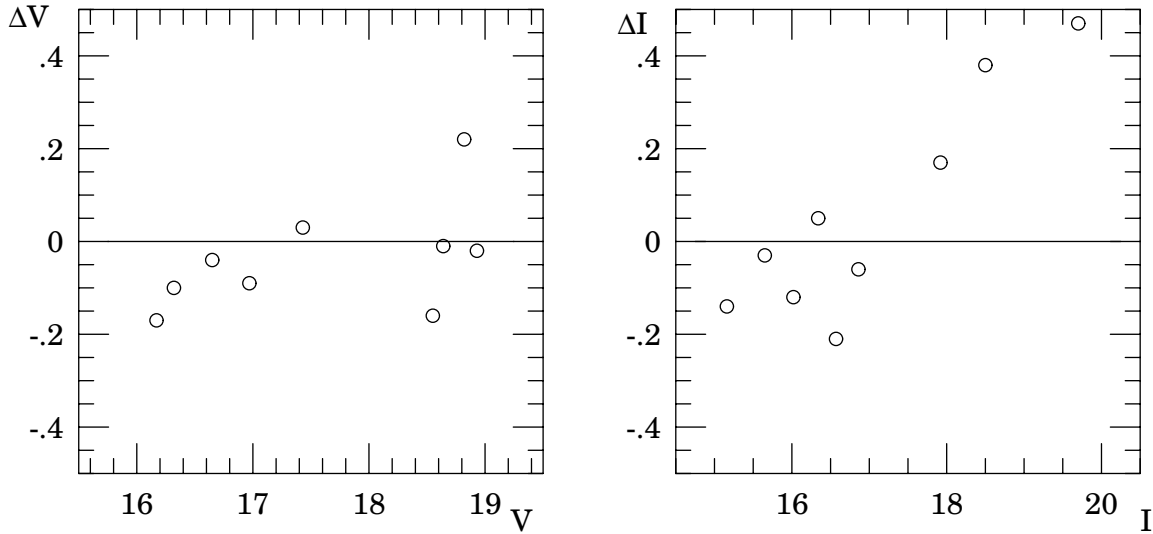}
\caption[]{A comparison of $V$ and $I$ magnitudes for objects in M31 and
M33 found in two fields} 
\label{cmpI}
\end{figure}

Taking advantage of the fact that there was some overlap between
subsequent fields the magnitudes of clusters caught on two frames were
compared. Such a comparison was to provide information on random errors
associated with the implemented procedure of determining magnitudes. The
results of the comparison are shown graphically in Figure~\ref{cmpI}.

The $V$ magnitudes are self-consistent to a sufficient degree over a wide
range. The differences between the $I$ magnitudes of the same objects show
a growing tendency with decreasing magnitude. A possible explanation of
this phenomenon is that for each field different star finding thresholds
were used. Thus the number of stars subtracted from the regions surrounding
those objects was not the same in each case. A closer look at two such
objects, 39 and 50 in the M31, revealed that the magnitudes measured in
field B were lower than the values obtained for field C and, as expected,
more stars were subtracted from frame B than frame C (23 stars in field B 
against 17 in field C for the first object and 15 versus 10 for the
second). This effect was much weaker in the $V$ band since there were far
less objects near the resolution limit visible in those frames. In any case
the $I$ magnitudes below 17th mag should be treated with reserve.

The measurement errors were estimated at $\sigma(B)=0.05, \sigma(V)=0.9,
\sigma(I)=0.18$. These values were obtained by finding the average
difference between the globular cluster magnitudes measured on different
frames. 

\label{cmp}

\subsubsection{Comparison with other catalogues}

\begin{figure}[hp]
\vspace{12.cm}
%\hspace{-3.5cm}
\includegraphics{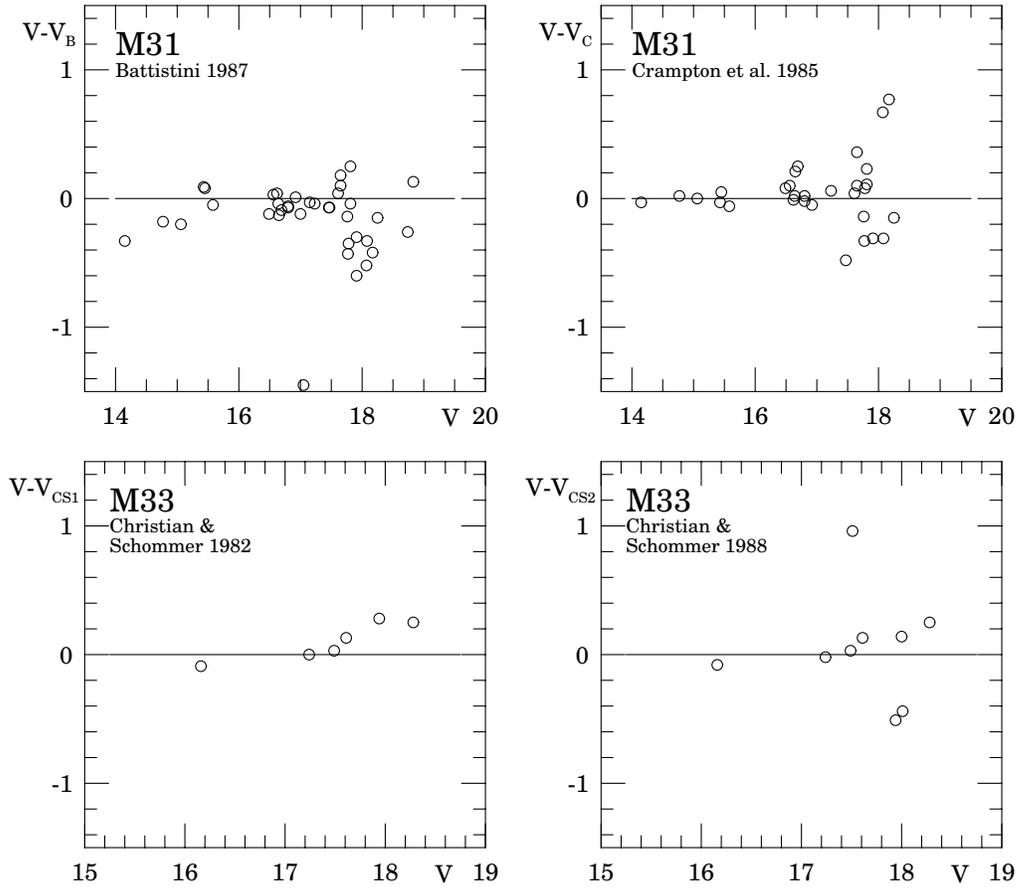}
\caption[]{A comparison of our $V$ magnitudes with data from other sources}
\label{cmpo}
\end{figure}

Magnitudes for the 38 already known globular clusters within the M31
fields were taken from the two following sources: Battistini 1987
($V_B$); Crampton et al.~1985 ($V_C$).  In the M33 16 previously
discovered objects were identified as globular clusters, and the
magnitudes of some of those objects were found in Christian \&
Schommer 1982 ($V_{CS1}$) and Christian \& Schommer 1988
($V_{CS2}$). The results of the comparison are showed in
Figure~\ref{cmpo}.

The magnitudes of globular clusters in the M31 measured here showed
good consistency with the values determined by others. The above
comparison gave no indication for the existence of meaningful
systematic errors in our measurements.

In the case of M33 there were relatively few objects available for
comparison. Taking that into consideration and the fact that both
measurements were conducted by the same persons, probably using the
same method, no definite conclusions can be drawn from the
comparison. The magnitudes of the brighter objects are consistent with
our values. For the fainter objects there seems to be a tendency for
our magnitudes to be lower than those of Christian \&
Schommer. Another striking feature is that the three objects appearing
only in the second diagram have magnitudes that differ by $\sim0.5-1$
mag. from our values. Taking into consideration the results of a
similar comparison in M31 it seems plausible to assume that the errors
of our measurements contribute a smaller amount to those large
differences.

\subsection{Luminosity functions}

In order to obtain a qualitative comparison of the objects found in
M31 and M33 with the globular cluster system of the Milky Way,
luminosity functions were constructed for each of those galaxies.

Data on 143 Milky Way clusters was taken from Harris 1996. The
following distance moduli were used: $\mu_{0,M31}=24.47$ mag (Stanek
\& Garnavich 1998), $\mu_{0,M33}=24.63$ mag (Madore \& Freeman
1991). Only the clusters visible in our fields are included in the
histograms for M31 and M33. It should be noted that those fields cover
only parts of the galaxies, so the sample is not complete. Extinction
hasn't been accounted for. Three histograms were drawn for the M31 and
M33 clusters containing: a) all of the cluster candidates found in our
fields; b) A, B and C class clusters; c) A and B class clusters.

\begin{figure}[htbp]
\vspace{14.5cm}
\hspace{-4.25cm}
\includegraphics{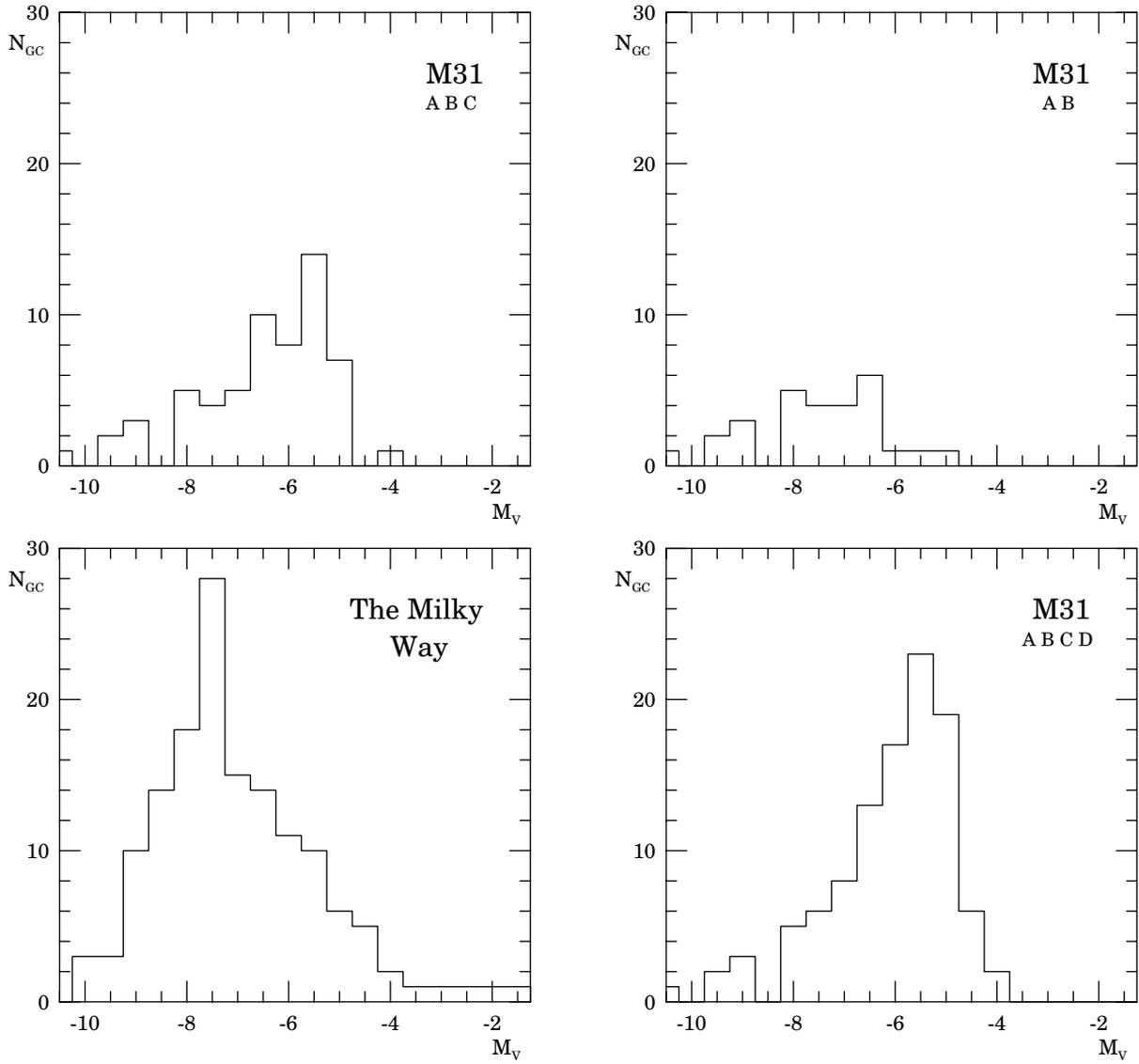}
\caption[]{A comparison of the luminosity functions in the Milky Way and M31}
\label{lf31}
\end{figure}

\begin{figure}[htbp]
\vspace{14.5cm}
\hspace{-4.25cm}
\includegraphics{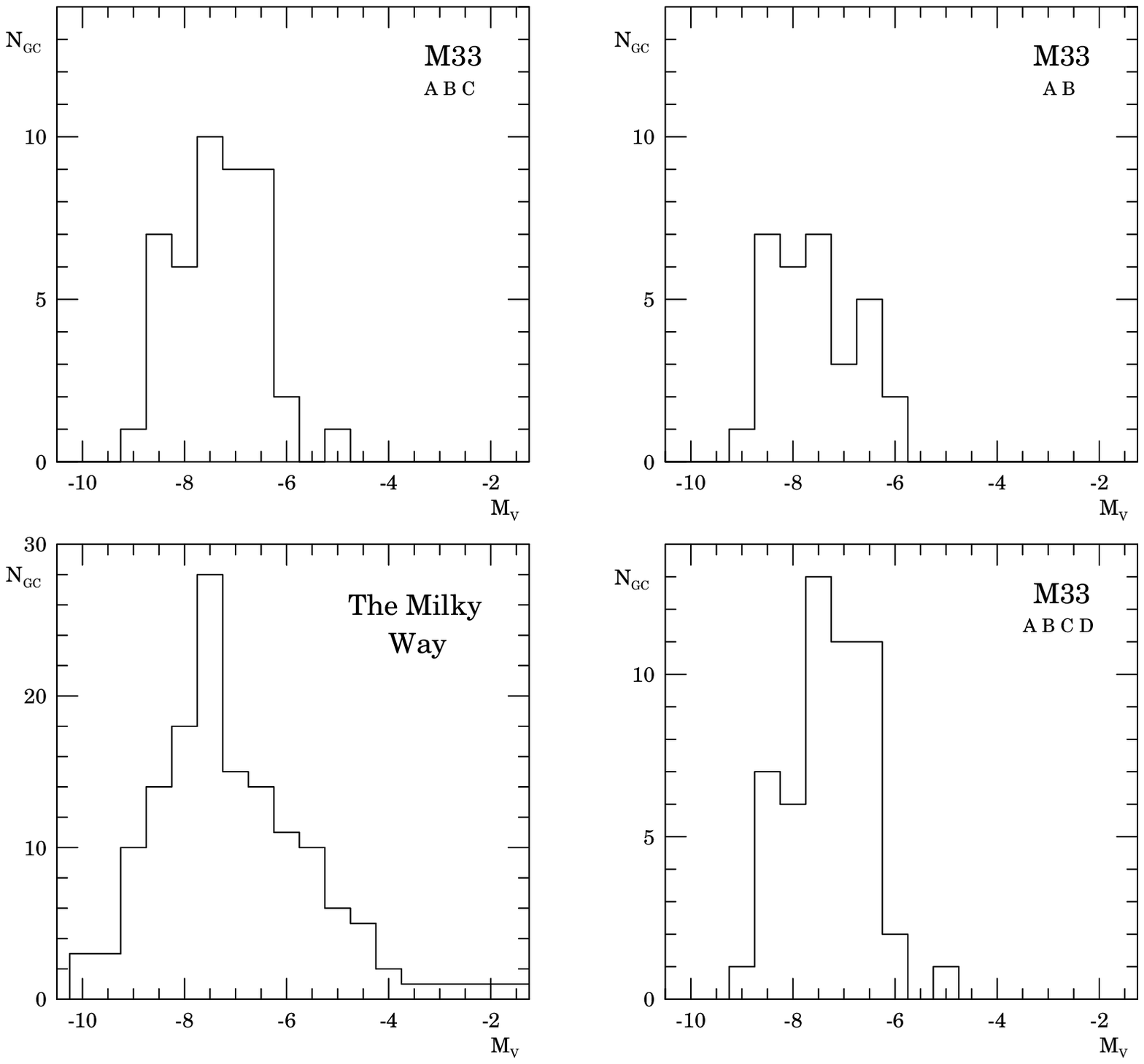}
\caption[]{A comparison of the luminosity functions in the Milky Way and M33}
\label{lf33}
\end{figure}

The luminosity functions are shown in Figs \ref{lf31} and \ref{lf33}.

The maximum of the luminosity function of the observed M31 disk
globular clusters appears to be shifted by about 2 magnitudes towards
lower brightness in comparison with the Milky Way. A comparison of the
halo globular cluster luminosity functions for the M31 and the Milky
Way shows that the difference between their turnover magnitudes is
$\sim0.3$ mag (Harris 1993), with the M31 clusters being
brighter. According to Gnedin \& Ostriker (1997) the influence of
dynamical effects on the cluster population is greater in the inner
parts of the galaxy. In particular the destruction mechanisms are
stronger in the central part. Their research indicates that low mass
(hence less luminous) clusters would have short lifetimes. In
accordance with the latter Gnedin (1997) found that the peak of the
inner globular cluster luminosity function is brighter by 0.8 mag than
for the outer clusters. Our luminosity function, constructed for the
previously discovered clusters and the new globular cluster
candidates, does not exhibit such behavior. Strong extinction due to
the large tilt of the galaxy could be a possible, although not very
likely, explanation. More research should be done in order to clarify
this situation. Higher resolution observational data should be
obtained for the newly discovered objects, most of them classified as
C or D, in order to verify their true nature. Not much can be said
about the shape of the luminosity function for A and B class clusters
in M31 due to the possibility of large statistical fluctuations within
such a small sample of objects. In particular the maximum of that
luminosity function is shifted with respect to the other two. Another
noteworthy feature is the brightness cutoff at \mbox{$V \sim -10$
mag.} seen in both the M31 and the Milky Way.

The luminosity function of the globulars in M33 has a maximum at the
same brightness as is observed in our Galaxy. M33 is seen face on and
the extinction is much weaker than in the case of M31.

\subsection{Color-magnitude diagrams}

\begin{figure}[htbp]
\vspace{12.5cm}
\includegraphics{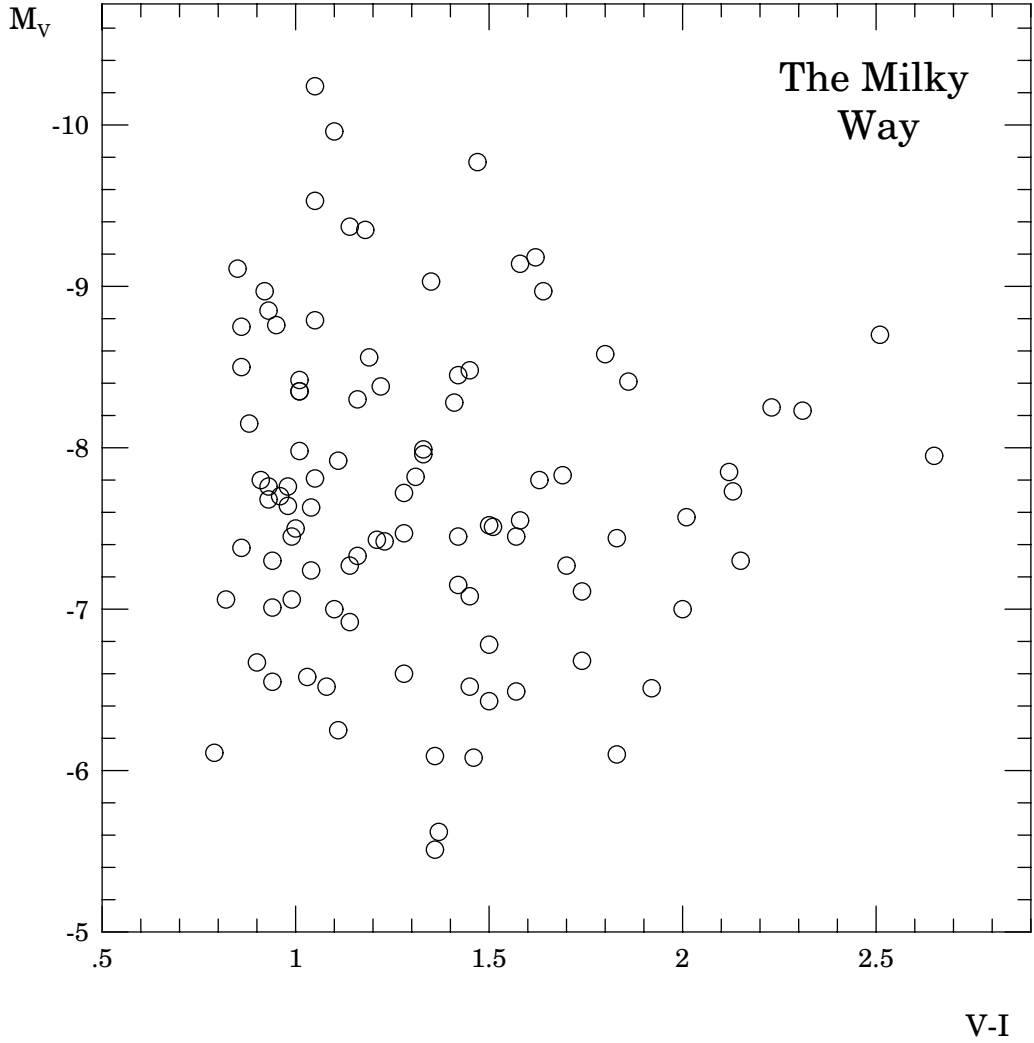} 
\caption[]{The $V/V-I$ color-magnitude diagram for the Milky Way globular
clusters} 
\label{mwVIcmd}
\end{figure}

\begin{figure}[htbp]
\vspace{12.5cm}
\includegraphics{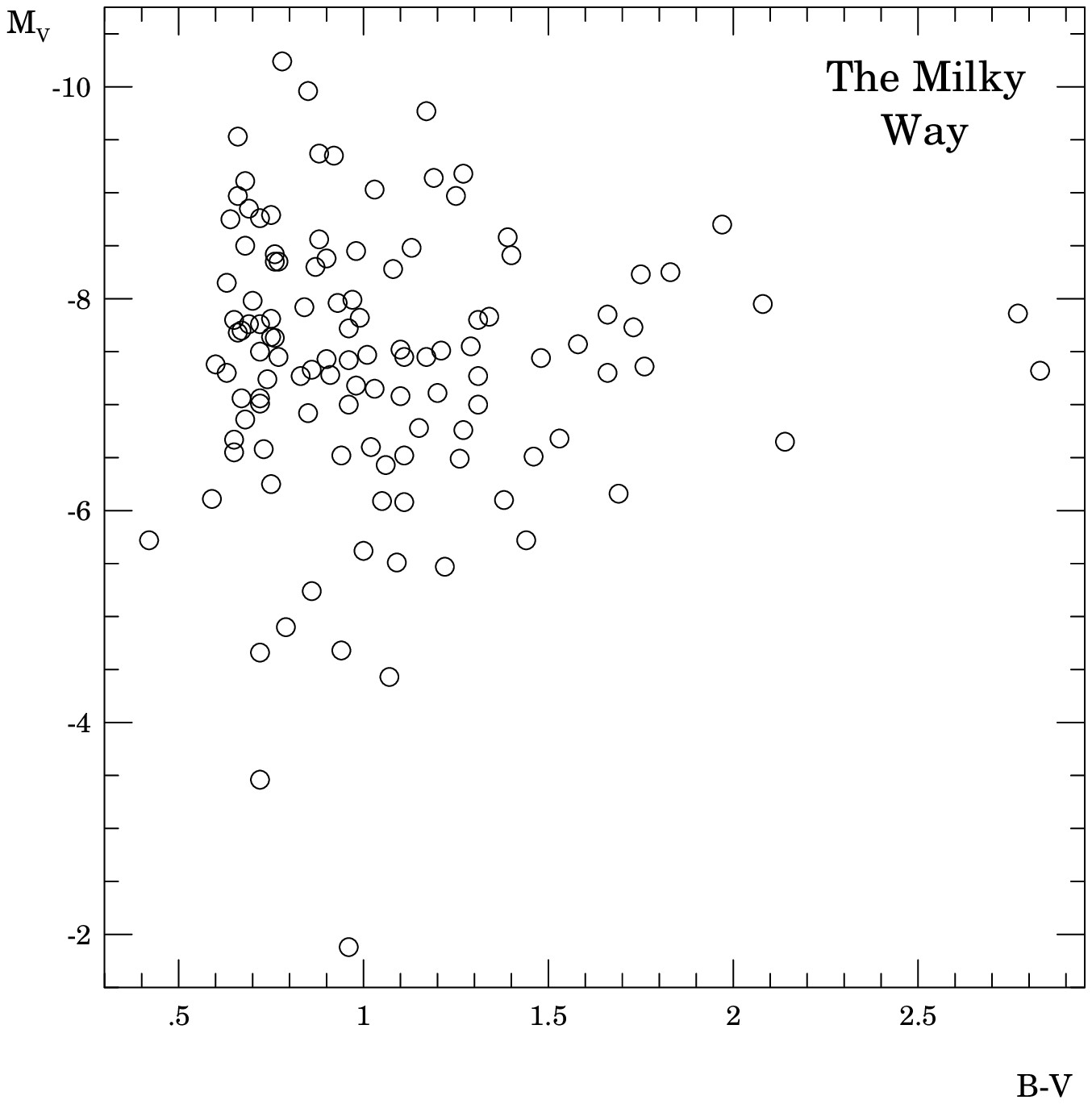}
\caption[]{The $V/B-V$ color-magnitude diagram for the Milky Way globular
clusters. } 
\label{mwBVcmd}
\end{figure}

\begin{figure}[htbp]
\vspace{12.5cm}
\includegraphics{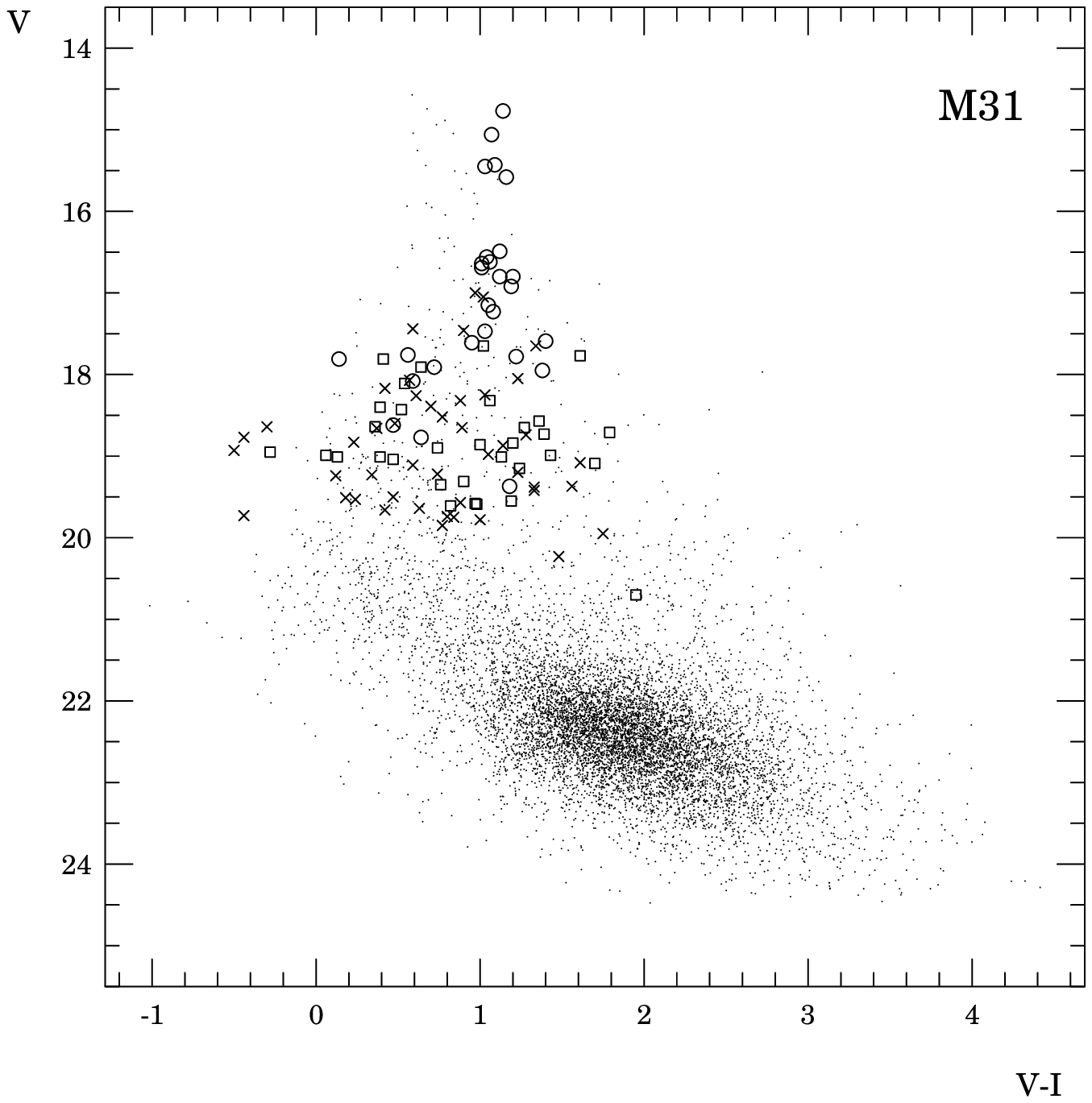}
\caption[]{The $V/V-I$ color-magnitude diagram for the M31 globular
cluster candidates. Class A and B clusters are denoted by circles
($\circ$), class C by boxes ($\Box$) and class D by crosses ($\times$).}
\label{31VIcmd}
\end{figure}

\begin{figure}[htbp]
\vspace{12.5cm}
\includegraphics{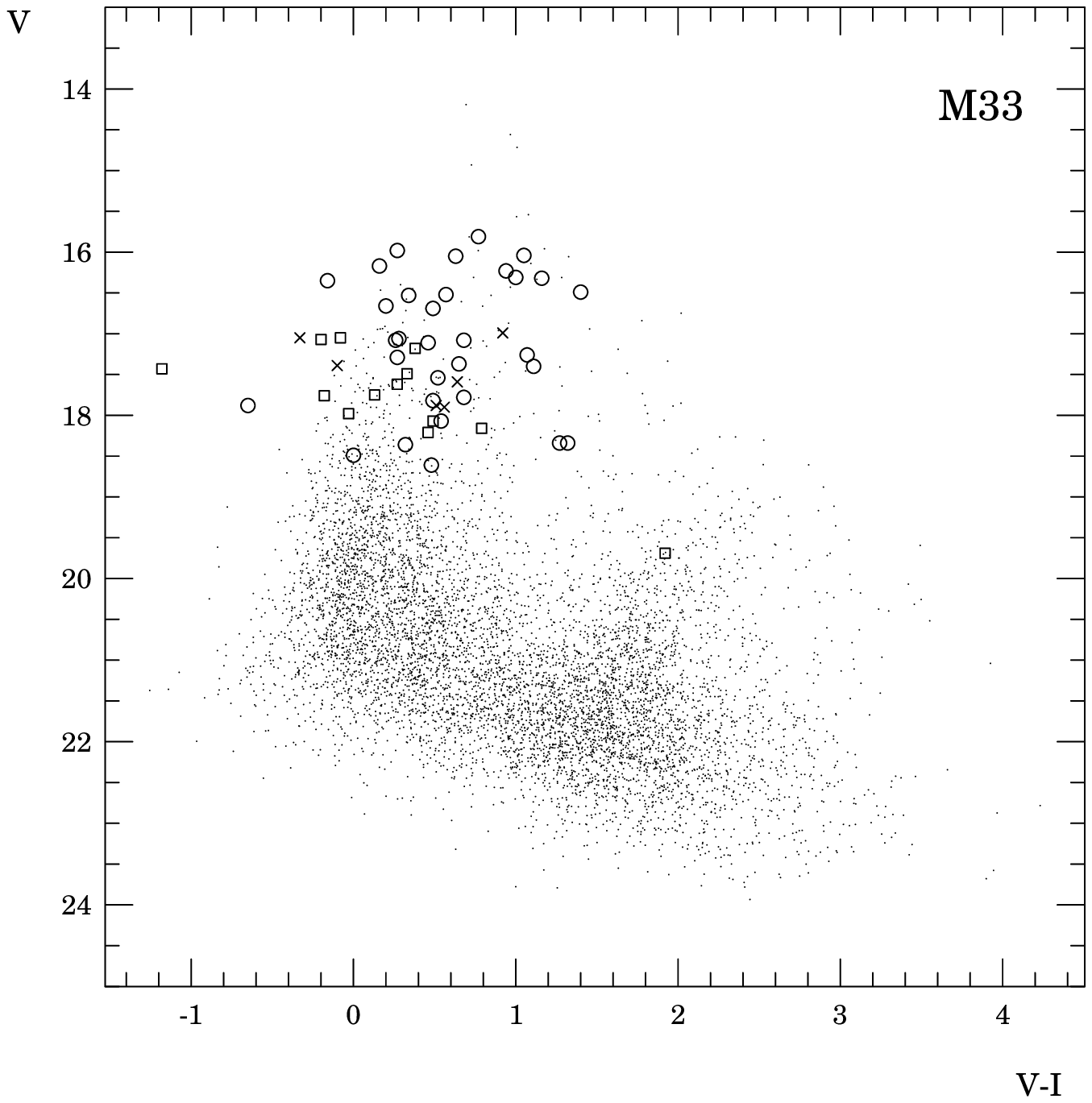}
\caption[]{The $V/V-I$ color-magnitude diagram for the M33 globular
cluster candidates.  Class A and B clusters are denoted by circles 
($\circ$), class C by boxes ($\Box$) and class D by crosses ($\times$).}
\label{33VIcmd}
\end{figure}

\begin{figure}[htbp]
\vspace{12.5cm}
\includegraphics{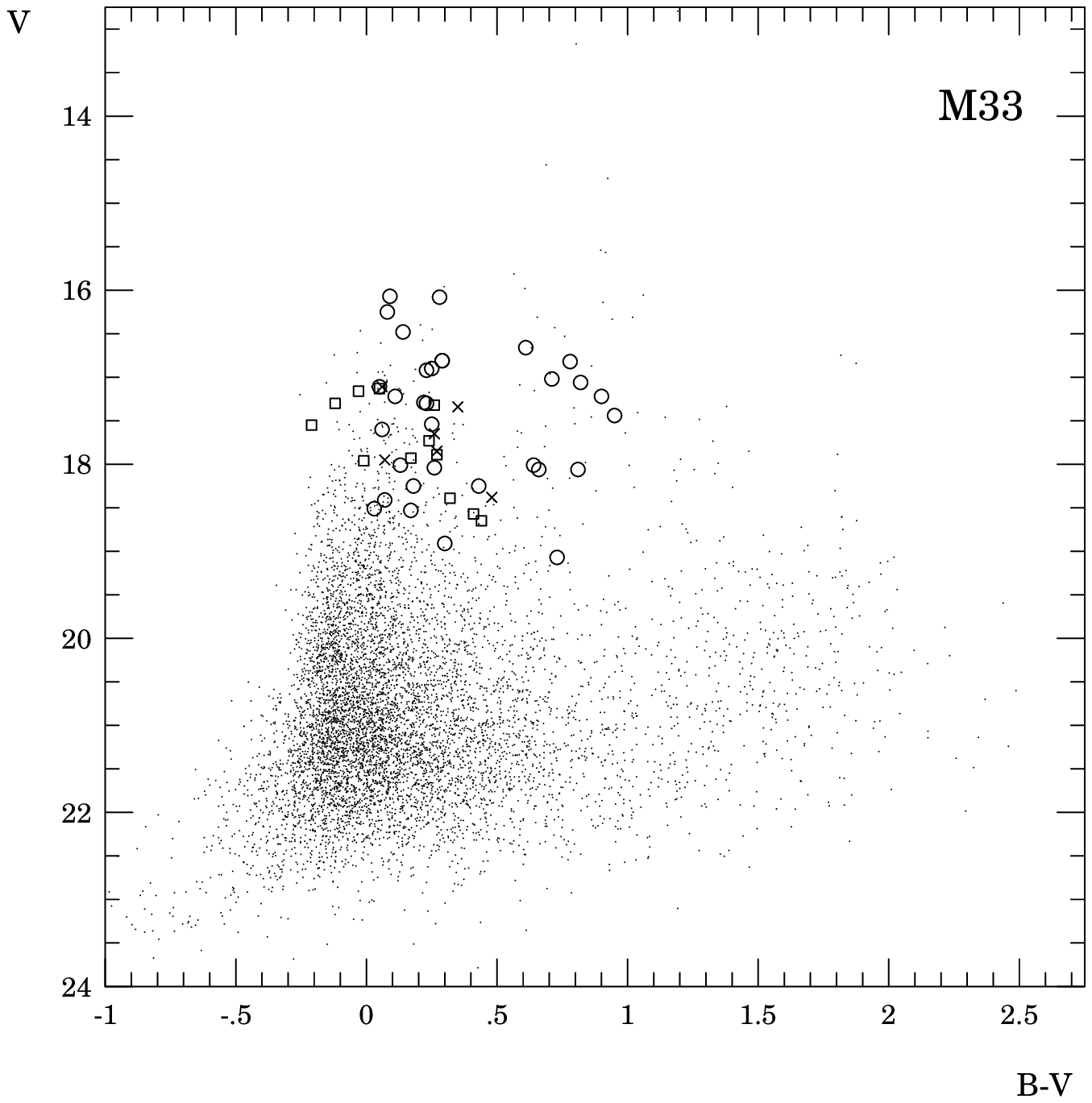}
\caption[]{The $V/B-V$ color-magnitude diagram for the M33 globular
cluster candidates.  Class A and B clusters are denoted by circles 
($\circ$), class C by boxes ($\Box$) and class D by crosses ($\times$).}
\label{33BVcmd}
\end{figure}

The color-magnitude diagrams for the globular clusters are shown in
Figs \ref{mwVIcmd} - \ref{33BVcmd}. Objects belonging to different
classes are marked with different symbols. Class A and B clusters are
denoted by circles ($\circ$), class C by boxes ($\Box$) and class D by
crosses ($\times$). The stars shown in the background come from fields
M31A and M33C. The photometric data for the Milky Way globulars was taken
from Harris (1996).

The color-magnitude diagrams for the Milky Way's globular clusters exhibit
sharp color cutoffs at \mbox{$V-I \sim 0.8$} and \mbox{$B-V \sim 0.6$},
with all of the clusters being located to the right of those lines.

In the $V/V-I$ diagram for M31 there is a sharp cutoff at $V-I\sim1$ for
objects brighter than $\sim 17$ mag, a value very similar to the one
observed for the Milky Way's globulars. A situation like that is not seen
for fainter objects. This may result from the lower accuracy of $I$
magnitude values below 17 mag, as was mentioned in section \ref{cmpI}. 

The color-magnitude diagrams for M33 show color cutoffs at \mbox{$V-I
\sim 0 - 0.2$} and \mbox{$B-V \sim 0.1$}. The globular clusters in the M33
are, on the average, bluer than their Milky Way counterparts. In the
$V/V-I$ diagram a B class object (A13) is seen far to the left of the
region occupied by other globulars. The faintness of the object in the $I$
band (18.54 mag.) is a probable cause of the error in the $V-I$ color. 

\subsection{The color-color diagram}

\begin{figure}[htbp]
\vspace{14.5cm}
\includegraphics{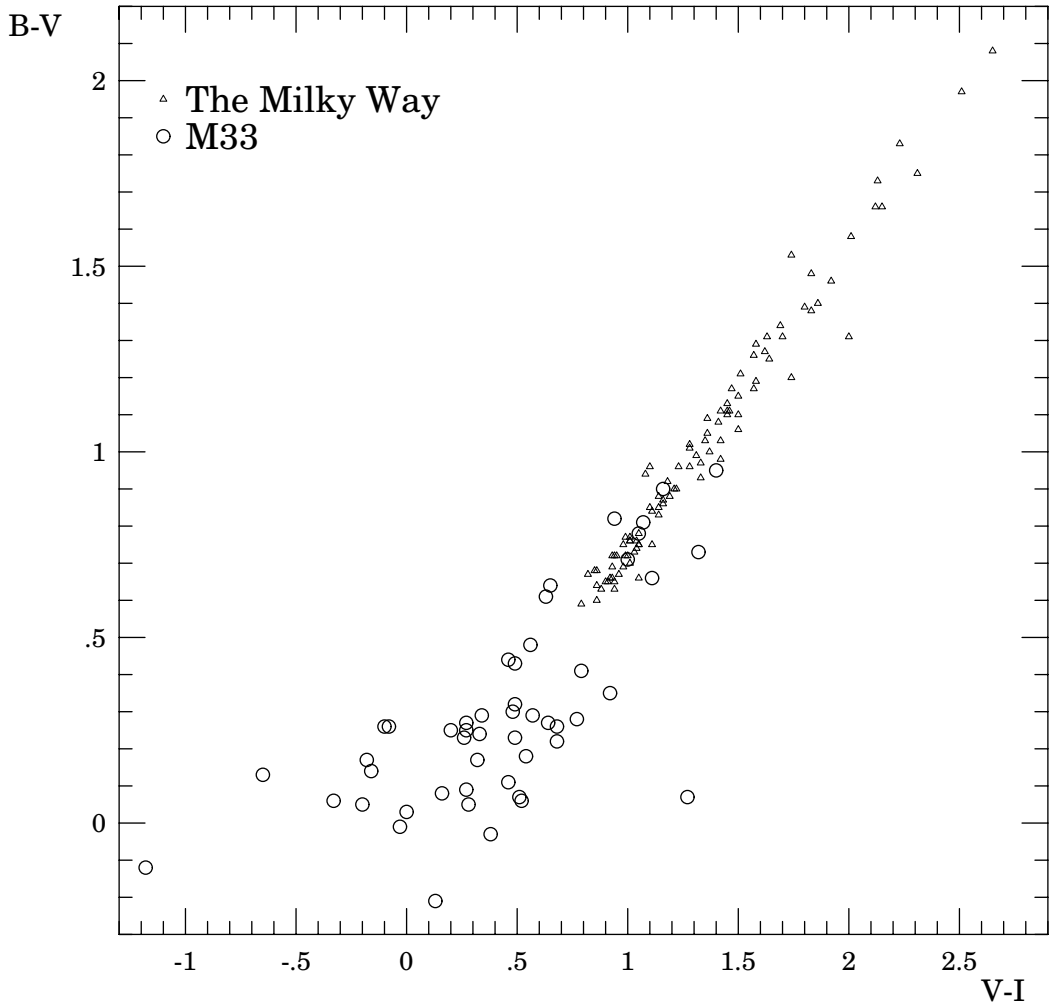}
\caption[]{The $V-I/B-V$ color-color diagram for the Milky Way and M33} 
\label{BVI}
\end{figure}

The $V-I/B-V$ diagram for the globular clusters in the Milky Way and
the M33 is shown in Fig. \ref{BVI}. The M33 globulars are shown as
open circles and the ones in the Milky Way by triangles. Data for the
Milky Way was taken from Harris (1996).

On the $V-I/B-V$ plane the Milky Way globular clusters are located in
a nearly straight line extending approximately between the points
$(0.8,0.6)$ and $(2.5,2.1)$. In the case of M33 the clusters also
appear to form a roughly straight line of a similar slope, shifted
towards lower color values.

\subsection{A double globular cluster candidate}

An object that could possibly be a double globular cluster was found
in field A in the M31 galaxy. Both components were assigned a C class
because of the difficulty in estimating the degree of contribution of
the other object to the widening of the radial profile. The magnitudes
were determined separately for each of the components by masking out
the other with a value of the median computed in a box of 101 pixels
on a side.  
%A separate map has been included for this object (Figure~\ref{GC2m}).

%\begin{figure}[htbp]
%\vspace{15cm}
%\special{psfile=gc2m.ps hoffset=0 voffset=0 vscale=100 hscale=100 angle=0}
%\caption[]{A double globular cluster candidate in the M31 (88 and 89).
%North is up and east is to the left.} 
%\label{GC2m}
%\end{figure}

\acknowledgements{We thank Pauline Barmby and John Huchra for their 
useful comments. JK was supported by NSF grant AST-9528096 to Bohdan 
Paczy\'nski and by the Polish KBN grant 2P03D011.12. KZS was supported 
by the Harvard-Smithsonian Center for Astrophysics Fellowship.}

%\subsection{Tables}

\begin{table}[htb]
\begin{center}
\caption[]{Globular cluster candidates in the M31 galaxy\\
{\scriptsize
Description:\\
G   -- Sargent et al. 1977\\
Bo  -- Battistini et al. 1987\\
Vet -- Vetesnik 1962\\
DAO -- Crampton et al. 1985\\
c   -- object class on a scale from A to D (see p. \pageref{klasy} for more
details)\\ 
f -- the field in which the object is located
 }
}
{\scriptsize
\begin{tabular}{lrrlllrlllll}
\hline\hline
ID&RA (2000)&DEC (2000)&V&I&V-I&c&f&G&Bo&Vet&DAO\\
\hline                                                                          
  1& 0:43:45.08&  41:18:12.2&  19.38 & 18.05 & ~1.33 & D&D&     &      &       &  \\          
  2& 0:43:49.07&  41:14:15.6&  18.32 & 17.44 & ~0.88 & D&D&     &      &       &  \\ 
  3& 0:43:54.23&  41:14:11.6&  18.77 & 19.21 & -0.44 & D&D&     &      &       &  \\ 
  4& 0:43:57.71&  41:13:52.5&  17.44 & 16.85 & ~0.59 & D&D&     &      &       &  \\ 
  5& 0:43:57.96&  41:21:33.2&  19.73 & 20.17 & -0.44 & D&C&     &      &       &  \\ 
  6& 0:43:58.20&  41:24:38.5&  15.43 & 14.34 & ~1.09 & A&C& G256& Bo205&   V102&  \\ 
  7& 0:43:58.64&  41:30:17.9&  15.06 & 14.00 & ~1.07 & A&C& G257& Bo206&   V106&  \\ 
  8& 0:44: 0.02&  41:23:11.6&  17.78 & 16.56 & ~1.22 & A&C& G259& Bo208&   V276&  \\ 
  9& 0:44: 0.81&  41:17:12.5&  17.95 & 16.57 & ~1.38 & A&D&     &      &       &  \\ 
 10& 0:44: 2.59&  41:25:26.5&  16.64 & 15.63 & ~1.01 & A&C& G261& Bo209&   V284&  \\ 
 11& 0:44: 2.71&  41:14:24.5&  17.81 & 17.40 & ~0.41 & C&D&     & Bo210&       &  \\
 12& 0:44: 2.81&  41:21:40.2&  18.77 & 18.13 & ~0.64 & B&C&     &      &       &  \\ 
 13& 0:44: 2.83&  41:20: 4.3&  16.65 & 15.65 & ~1.00 & A&C& G262& Bo211&   V98 &  \\ 
   &           &            &  16.69 & 15.68 & ~1.01 & A&D& G262& Bo211&   V98 &  \\ 
 14& 0:44: 3.47&  41:30:38.7&  16.92 & 15.73 & ~1.19 & A&C& G264& Bo213&   V292&  \\ 
 15& 0:44: 3.94&  41:26:18.5&  17.61 & 16.65 & ~0.95 & A&C& G265& Bo214&   V286&  \\ 
 16& 0:44: 8.02&  41:23:54.2&  19.37 & 17.81 & ~1.56 & D&C&     &      &       &  \\ 
 17& 0:44:10.62&  41:23:50.9&  16.49 & 15.37 & ~1.12 & A&C& G269& Bo217&   V281&  \\ 
 18& 0:44:11.22&  41:25:22.2&  17.59 & 16.19 & ~1.40 & B&C&     &      &   V285&  \\ 
 19& 0:44:11.66&  41:23:53.8&  18.62 & 18.15:& ~0.47:& B&C&     &      &       &  \\ 
 20& 0:44:13.90&  41:22:18.7&  18.93 & 19.43 & -0.50 & D&C&     &      &       &  \\ 
 21& 0:44:14.26&  41:19:19.6&  14.77 & 13.64 & ~1.14 & A&D& G272& Bo218&   V101&  \\ 
 22& 0:44:16.87&  41:14:15.9&  18.74 & 17.46 & ~1.28 & D&D&     & Bo277&       &  \\
 23& 0:44:18.86&  41:21: 9.8&  18.65 & 17.75 & ~0.89 & D&D&     &      &       &  \\
   &           &            &  18.64 & 17.92 & ~0.72 & D&C&     &      &       &  \\
 24& 0:44:19.34&  41:30:35.3&  16.62 & 15.55 & ~1.06 & B&C& G275& Bo220&   V114&  \\ 
 25& 0:44:19.64&  41:24: 8.5&  18.64 & 18.93 & -0.30 & D&C&     &      &       &  \\ 
 26& 0:44:20.22&  41:27:20.0&  20.23 & 18.75 & ~1.48 & D&C&     &      &       &  \\ 
 27& 0:44:21.16&  41:19: 9.9&  18.65 & 17.38 & ~1.27 & C&D&     &      &       &  \\ 
 28& 0:44:22.05&  41:33:40.1&  18.66 & 18.29 & ~0.37 & D&B&     &      &       &  \\ 
 29& 0:44:23.05&  41:33: 6.4&  16.80 & 15.68 & ~1.12 & A&B& G276& Bo221&   V117&  \\
 30& 0:44:23.32&  41:35: 4.1&  18.83 & 18.60 & ~0.23 & D&B&     & Bo278&       &  \\
 31& 0:44:24.28&  41:33:58.5&  18.90 & 18.15 & ~0.74 & C&B&     &      &       &  \\ 
 32& 0:44:25.33&  41:14:11.9&  17.65:& 16.63 & ~1.02:& C&D& G277& Bo222&   V96 &  \\
 33& 0:44:26.51&  41:38:57.5&  18.32 & 17.26 & ~1.06 & C&B&	&      &       &  \\
 34& 0:44:27.03&  41:34:36.7&  17.81 & 17.67 & ~0.14 & B&B& G278& Bo223&       &  \\
 35& 0:44:27.06&  41:28:50.2&  15.45 & 14.42 & ~1.03 & A&C& G279& Bo224&   V112&  \\
 36& 0:44:27.48&  41:11:34.0&  17.46 & 16.55 & ~0.90 & D&D&     & Bo473&       &  \\
 37& 0:44:29.49&  41:21:35.7&  14.15 & ------& ~-----& A&C& G280& Bo225&   V282&  \\
 38& 0:44:30.51&  41:10:58.9&  17.65 & 16.32 & ~1.34 & D&D&     & Bo226&       &  \\
 39& 0:44:31.30&  41:30: 4.6&  18.93 & 19.70 & -0.77 & C&B&	&      &	& \\
   &           &            &  18.95 & 19.23 & -0.28 & C&C&     &      &       &  \\
 40& 0:44:31.49&  41:27:55.2&  19.59 & 18.61 & ~0.98 & C&C&     &      &       &  \\
 41& 0:44:33.89&  41:38:28.1&  16.56 & 15.53 & ~1.04 & A&B& G282& Bo229&   V120&  \\
 42& 0:44:33.90&  41:21: 3.2&  18.64 & 18.27 & ~0.36 & C&C&     &      &       &  \\
 43& 0:44:34.29&  41:23:11.6&  19.64 & 19.01:& ~0.63:& D&C&     &      &       &  \\
 44& 0:44:34.43&  41:24: 9.3&  19.24 & 19.12 & ~0.12 & D&C&     &      &       &  \\
 45& 0:44:36.39&  41:35:32.9&  18.73 & 17.34 & ~1.39 & C&B&     &      &       &  \\
 46& 0:44:36.64&  41:27:13.7&  19.01 & 17.88 & ~1.13 & C&C&     &      &       &  \\
 47& 0:44:37.85&  41:28:52.1&  18.84 & 17.64 & ~1.20 & C&C&     &      &       &  \\
\hline
\end{tabular}
}
\end{center}
\end{table}
\clearpage
\begin{table}[htb]
\begin{center}
\caption[]{Globular cluster candidates in the M31 galaxy (continued)}
{\scriptsize
\begin{tabular}{rrrlllrrrrrr}
\hline
ID&RA(2000)&DEC(2000)&V&I&V-I&c&f&G&Bo&Vet&DAO\\
\hline
 48& 0:44:38.60&  41:27:47.4&  17.23 & 16.15 &  ~1.08 & A&C& G286& Bo231&       &  \\
 49& 0:44:39.70&  41:24:28.2&  18.26 & 17.65 &  ~0.61 & D&C&	&     &	     &	  \\
 50& 0:44:40.59&  41:30: 6.0&  18.60:& 18.12:&  ~0.48:& D&C&	&     &	     &	  \\
   &           &            &  18.82:& 18.50 &  ~0.32:& D&B&	&     &	     &	  \\
 51& 0:44:46.37&  41:29:17.7&  16.80 & 15.60 &  ~1.20 & A&C& G290& Bo234&   V297&  \\
 52& 0:44:56.15&  41:41:35.7&  19.74 & 18.93 &  ~0.80 & D&A&  &      &       &     \\
 53& 0:44:57.22&  41:48: 2.3&  19.09 & 17.39 &  ~1.70 & C&A&  &      &       &     \\
 54& 0:44:59.13&  41:42:25.2&  19.55 & 18.36 &  ~1.19 & C&A&  &      &       &     \\
 55& 0:44:59.61&  41:33:39.3&  18.99 & 17.56 &  ~1.43 & C&B&  &      &       &     \\
 56& 0:45: 1.54&  41:39: 4.4&  19.58 & 18.61 &  ~0.97 & C&A&  &      &       &     \\
 57& 0:45: 2.73&  41:47: 2.5&  19.95 & 18.20 &  ~1.75 & D&A&  &      &       &     \\
 58& 0:45: 3.31&  41:40: 5.6&  18.55 & 16.86 &  ~1.68 & C&A&  &     &	     &	  \\
   &           &            &  18.71 & 16.92 &  ~1.79 & C&B&  &     &	     &	  \\
 59& 0:45: 4.03&  41:46:20.7&  18.43 & 17.92 &  ~0.52 & C&A&  &      &       &     \\
 60& 0:45: 6.81&  41:38:57.7&  19.23 & 18.89 &  ~0.34 & D&A&  &      &       &     \\
 61& 0:45: 7.18&  41:40:31.4&  18.25 & 17.22 &  ~1.03 & D&A&  & Bo476&       &  D74\\
 62& 0:45: 7.61&  41:45:31.0&  19.22 & 18.48 &  ~0.74 & D&A&  &      &       &     \\
 63& 0:45: 8.30&  41:39:38.0&  18.17 & 17.75 &  ~0.42 & D&B&  & Bo477&       &     \\
 64& 0:45: 8.31&  41:39:38.0&  18.07 & 17.50 &  ~0.57 & D&A&  & Bo477&       &  D75\\
 65& 0:45: 9.89&  41:42:22.9&  19.01 & 18.88 &  ~0.13 & C&A&  &      &       &     \\
 66& 0:45:10.94&  41:40:23.0&  17.00 & 16.04 &  ~0.97 & D&B&  & Bo260&       &     \\
 67& 0:45:10.96&  41:40:23.0&  17.05 & 16.03 &  ~1.02 & D&A&  & Bo260&       &     \\
 68& 0:45:10.97&  41:38:56.3&  19.75 & 18.91:&  ~0.84:& D&A&  &      &       &     \\
 69& 0:45:11.27&  41:49:20.3&  19.20 & 17.97 &  ~1.23 & D&A&  &      &       &     \\
 70& 0:45:11.72&  41:40:20.1&  19.08 & 17.47 &  ~1.61 & D&A&  &      &       &     \\
 71& 0:45:13.75&  41:42:34.2&  19.53 & 19.29 &  ~0.24 & D&A&  &      &       &     \\
 72& 0:45:13.81&  41:42:26.1&  18.40 & 18.01 &  ~0.39 & C&A&  &      &       &     \\
 73& 0:45:15.13&  41:47:32.2&  19.37 & 18.18 &  ~1.18 & A&A&  &      &       &     \\
 74& 0:45:15.60&  41:35:17.4&  17.15 & 16.10 &  ~1.05 & A&B&  & Bo239&       &     \\
 75& 0:45:15.80&  41:44:49.6&  20.70 & 18.76 &  ~1.95 & C&A&  &      &       &     \\
 76& 0:45:16.07&  41:43:22.2&  18.99 & 18.93 &  ~0.06 & C&A&  &      &       &     \\
 77& 0:45:17.37&  41:39:32.9&  19.51:& 19.34 &  ~0.18:& D&A&  &      &       &     \\
 78& 0:45:17.72&  41:41:52.6&  19.50 & 19.03 &  ~0.47 & D&A&  &      &       &     \\
 79& 0:45:17.74&  41:40:58.0&  19.01 & 18.61 &  ~0.39 & C&A&  &      &       &     \\
 80& 0:45:19.57&  41:48:30.3&  18.52 & 17.74:&  ~0.77:& D&A&  &      &       &     \\
 81& 0:45:22.26&  41:47:57.1&  19.78 & 18.78 &  ~1.00 & D&A&     &      &     &    \\ 
 82& 0:45:26.23&  41:45:52.3&  19.04 & 18.57 &  ~0.47 & C&A&     &      &     &    \\ 
 83& 0:45:26.94&  41:45:43.5&  19.11 & 18.52:&  ~0.59:& D&A&     &      &     &    \\ 
 84& 0:45:27.15&  41:43:45.2&  17.76 & 17.20:&  ~0.56:& B&A& G303& Bo371& V122&    \\ 
 85& 0:45:27.97&  41:42: 4.1&  19.57 & 18.70 &  ~0.88 & D&A&     &      &     &    \\ 
 86& 0:45:28.44&  41:49:29.1&  18.11 & 17.57 &  ~0.54 & C&A&     &      &     &    \\ 
 87& 0:45:31.98&  41:49:32.2&  18.05 & 16.81 &  ~1.23 & D&A&     &      &    &     \\ 
 88& 0:45:32.47&  41:43:33.8&  19.31:& 18.41:&  ~0.90:& C&A&     &      &    &     \\ 
 89& 0:45:32.53&  41:43:31.3&  18.57:& 17.22:&  ~1.36:& C&A&     &      &    &     \\ 
 90& 0:45:32.94&  41:48:22.8&  19.85 & 19.08 &  ~0.77 & D&A&     &      &    &     \\ 
 91& 0:45:33.12&  41:42:19.8&  18.86 & 17.87 &  ~1.00 & C&A&     &      &    &     \\ 
 92& 0:45:35.60&  41:45:18.4&  18.98 & 17.92:&  ~1.05:& D&A&     &      &    &     \\ 
 93& 0:45:37.44&  41:40:11.2&  19.35 & 18.59 &  ~0.76 & C&A&     &      &    &     \\ 
 94& 0:45:39.74&  41:46:34.6&  19.61 & 18.78 &  ~0.82 & C&A&     &      &    &     \\ 
 95& 0:45:39.82&  41:44:41.8&  19.15 & 17.91 &  ~1.24 & C&A&     &      &    &     \\ 
 96& 0:45:41.89&  41:45:33.5&  15.58 & 14.41 &  ~1.16 & A&A& G305& Bo373&V125&     \\ 
 97& 0:45:44.12&  41:40: 5.6&  17.77:& 16.16 &  ~1.61:& C&A&     & Bo268&    &  D82\\ 
 98& 0:45:44.54&  41:41:54.6&  18.08 & 17.49:&  ~0.59:& A&A& G306& Bo374&    &     \\ 
 99& 0:45:45.58&  41:45:52.9&  17.91 & 17.20:&  ~0.72:& B&A&     & Bo480&V127&     \\ 
100& 0:45:45.58&  41:39:42.5&  17.47 & 16.44:&  ~1.03:& A&A& G307& Bo375&    &     \\
101& 0:45:46.25&  41:48:20.3&  19.42 & 18.09 &  ~1.33 & D&A&    &   &     &  	  \\
102& 0:45:46.75&  41:45:22.7&  19.66 & 19.24:&  ~0.42:& D&A&    &    &     & 	  \\
103& 0:45:48.44&  41:42:40.0&  17.91 & 17.27:&  ~0.64:& C&A& G309& Bo376&V124&     \\ 
104& 0:45:48.82&  41:48:19.9&  18.39 & 17.68:&  ~0.70:& D&A&    &   &    &   	  \\
105& 0:45:49.66&  41:39:25.9&  18.87 & 17.73:&  ~1.14:& D&A&    &  &     &   	  \\
\hline	\hline	
\end{tabular}
}
\end{center}
\end{table}

\clearpage
\begin{table}[htb]
\begin{center}
{\scriptsize
\caption[]{Globular cluster candidates in the M33 galaxy\\
{\scriptsize
Description:\\
C\&S -- Christian \& Schommer 1982\\
c   -- object class on a scale from A to D (see p. \pageref{klasy} for more details)\\
f -- the field in which the object is located
 }
}
\begin{tabular}{lrrlllllrll}
\hline\hline
ID&RA (2000)&DEC (2000)&B&V&I&B-V&V-I&c&f&C\&S\\
\hline                                                                         
  1& 1:32:56.11& 30:38:25.42&  17.60 & 17.54 & 17.02 & ~0.06 & ~0.52 & A&C&	\\
  2& 1:33: 6.43& 30:37:35.55&  22.19 & 19.69 & 17.77 & ~2.50 & ~1.92 & C&C&     \\
  3& 1:33:20.42& 30:40:23.33&  17.29 & 17.08 & 16.40 & ~0.22 & ~0.68 & A&C&     \\
  4& 1:33:23.12& 30:33: 0.70&  17.54 & 17.29 & 17.02 & ~0.25 & ~0.27 & A&C& H32 \\
  5& 1:33:25.64& 30:29:56.98&  19.07 & 18.34 & 17.02 & ~0.73 & ~1.32 & A&C&     \\
  6& 1:33:26.01& 30:36:24.23&  18.04 & 17.78 & 17.10 & ~0.26 & ~0.68 & B&C&     \\
  7& 1:33:27.99& 30:32:43.23&  18.53 & 18.36 & 18.04 & ~0.17 & ~0.32 & A&C&     \\
  8& 1:33:29.51& 30:30: 2.17&  18.25:& 18.07:& 17.53:& ~0.18:& ~0.54:& A&C&     \\
  9& 1:33:31.01& 30:36:52.47&  18.41 & 18.34 & 17.06 & ~0.07 & ~1.27 & A&C&     \\
 10& 1:33:37.02& 30:37:12.01&  18.51:& 18.49:& 18.49:& ~0.03:& ~0.00:& B&C&     \\
 11& 1:33:37.27& 30:34:14.01&  17.11:& 17.06 & 16.78:& ~0.05:& ~0.28:& A&C& U111\\
   & 1:33:37.27& 30:34:14.24&  17.18 & 16.97:& 16.57:& ~0.21:& ~0.40:& B&B& U111\\
 12& 1:33:38.02& 30:38: 2.17&  18.06 & 17.40 & 16.29 & ~0.66 & ~1.11 & B&C&     \\ 
   & 1:33:38.09& 30:38: 3.49&  18.10:& 17.43 & 16.34 & ~0.68:& ~1.09 & C&B&     \\
 13& 1:33:38.05& 30:33: 5.49&  17.89:& 17.62:& 17.35:& ~0.27:& ~0.27:& C&B&     \\
 14& 1:33:39.66& 30:31: 8.97&  16.48 & 16.35 & 16.50 & ~0.14 & -0.16 & B&B&     \\
 15& 1:33:40.05& 30:38:27.62&  16.08 & 15.81:& 15.03:& ~0.28:& ~0.77:& B&B&     \\
 16& 1:33:40.40& 30:43:57.68&  17.22 & 17.11 & 16.65 & ~0.11 & ~0.46 & B&A&     \\
 17& 1:33:41.61& 30:41:42.81&  17.34 & 16.99 & 16.07 & ~0.35 & ~0.92 & D&A&     \\
 18& 1:33:42.99& 30:42:52.60&  17.13:& 17.07:& 17.27 & ~0.05:& -0.20:& C&A&     \\
 19& 1:33:43.86& 30:32:10.32&  17.55 & 17.75 & 17.63 & -0.21 & ~0.13 & C&B&     \\
 20& 1:33:44.61& 30:37:53.62&  17.30 & 17.43 & 18.60 & -0.12 & -1.18 & C&B& U94 \\
 21& 1:33:45.06& 30:47:46.73&  16.82 & 16.04 & 14.99 & ~0.78 & ~1.05 & A&A& U49 \\
 22& 1:33:52.14& 30:29: 3.63&  18.06 & 17.26 & 16.18 & ~0.81 & ~1.07 & B&B& H38 \\
 23& 1:33:54.15& 30:33: 9.67&  17.16 & 17.18 & 16.80 & -0.03 & ~0.38 & C&B&     \\
 24& 1:33:55.18& 30:47:57.88&  16.81 & 16.53 & 16.19 & ~0.29 & ~0.34 & B&A&     \\
 25& 1:33:57.89& 30:35:31.89&  18.65:& 18.21:& 17.75:& ~0.44:& ~0.46:& C&B&     \\
 26& 1:33:58.05& 30:45:44.93&  17.32:& 17.05:& 17.14:& ~0.26:& -0.08:& C&A& H14 \\
 27& 1:34: 0.29& 30:37:47.57&  16.66 & 16.05 & 15.42 & ~0.61 & ~0.63 & B&B&     \\
 28& 1:34: 1.61& 30:42:30.72&  18.01 & 17.88 & 18.54 & ~0.13 & -0.65 & B&A& U75 \\
 29& 1:34: 2.03& 30:39:37.35&  17.06 & 16.23 & 15.29 & ~0.82 & ~0.94 & A&A&     \\
 30& 1:34: 2.49& 30:38:40.98&  17.11 & 17.05 & 17.39 & ~0.06 & -0.33 & D&A&	\\
 31& 1:34: 2.51& 30:40:40.30&  17.44 & 16.49 & 15.10 & ~0.95 & ~1.40 & A&A&     \\
 32& 1:34: 2.83& 30:46:36.57&  17.85 & 17.59 & 16.95 & ~0.27 & ~0.64 & D&A&     \\
 33& 1:34: 2.94& 30:43:20.60&  17.02 & 16.31 & 15.31 & ~0.71 & ~1.00 & A&A&     \\
 34& 1:34: 4.35& 30:39:22.28&  17.95 & 17.88 & 17.37 & ~0.07 & ~0.51 & D&A&     \\
 35& 1:34: 7.22& 30:35:23.09&  17.96 & 17.98 & 18.01 & -0.01 & -0.03 & C&B&     \\
 36& 1:34: 8.06& 30:38:37.75&  17.22 & 16.32 & 15.16 & ~0.90 & ~1.16 & A&A&     \\ 
   & 1:34: 8.10& 30:38:38.61&  17.32 & 16.42 & 15.30 & ~0.90 & ~1.12 & A&B&     \\
 37& 1:34: 8.55& 30:39: 1.96&  16.25 & 16.17 & 16.02 & ~0.08 & ~0.16 & A&A&     \\ 
   & 1:34: 8.59& 30:39: 2.87&  16.35 & 16.34 & 16.14 & ~0.01 & ~0.20 & A&B&     \\
 38& 1:34: 9.00& 30:36:33.92&  17.93:& 17.76:& 17.94:& ~0.17:& -0.18:& C&B& H30 \\
 39& 1:34:10.13& 30:45:29.20&  17.65 & 17.39 & 17.49 & ~0.26 & -0.10 & D&A& U62 \\
 40& 1:34:10.69& 30:45:48.70&  16.07 & 15.98 & 15.71 & ~0.09 & ~0.27 & B&A&     \\
 41& 1:34:10.98& 30:40:29.75&  18.25 & 17.82 & 17.33 & ~0.43 & ~0.49 & B&A& U83 \\
 42& 1:34:11.40& 30:41:27.83&  18.57 & 18.16 & 17.37 & ~0.41 & ~0.79 & C&A& U78 \\
 43& 1:34:11.57& 30:34:52.43&  16.81 & 16.52 & 15.96 & ~0.29 & ~0.57 & A&B&     \\
 44& 1:34:14.20& 30:39:58.10&  18.38 & 17.90 & 17.34 & ~0.48 & ~0.56 & D&A& U82 \\
 45& 1:34:14.66& 30:32:35.05&  18.39:& 18.07:& 17.59:& ~0.32:& ~0.49:& C&B& H33 \\
 46& 1:34:15.09& 30:41:19.08&  17.73 & 17.49 & 17.17 & ~0.24 & ~0.33 & C&A& U79 \\
 47& 1:34:18.69& 30:31:37.67&  16.90 & 16.66 & 16.46 & ~0.25 & ~0.20 & B&B&     \\
 48& 1:34:19.47& 30:46:21.15&  16.92 & 16.69 & 16.20 & ~0.23 & ~0.49 & B&A&     \\
 49& 1:34:19.89& 30:36:12.68&  17.30 & 17.08 & 16.81 & ~0.23 & ~0.26 & A&B&     \\
 50& 1:34:20.21& 30:39:33.13&  18.91 & 18.61 & 18.13 & ~0.30 & ~0.48 & B&A& U91 \\
 51& 1:34:25.43& 30:41:28.48&  18.01:& 17.37:& 16.72:& ~0.64:& ~0.65:& B&A& H21 \\
\hline\hline
\label{t33a}
\end{tabular}
}
\end{center}	
\end{table}
\clearpage

%\subsection{Maps}

\begin{figure}[htbp]
\vspace{19.5cm}
\includegraphics{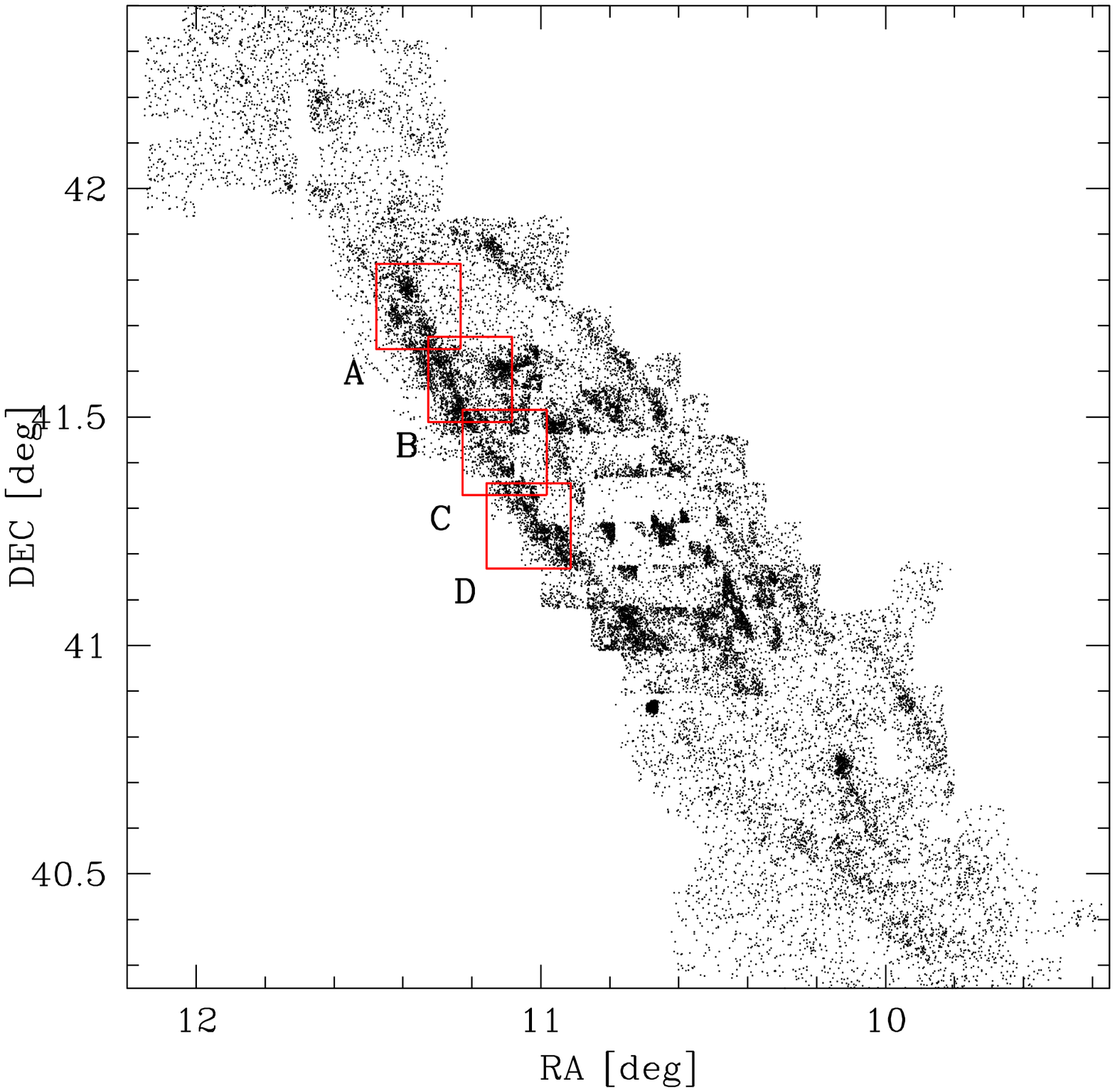} 
\vspace{-3.5cm}
\caption[]{The location of fields A - D in the M31 galaxy} 
\label{31f}
\end{figure}

\clearpage

\begin{figure}[htbp]
\vspace{19.75cm}
\includegraphics{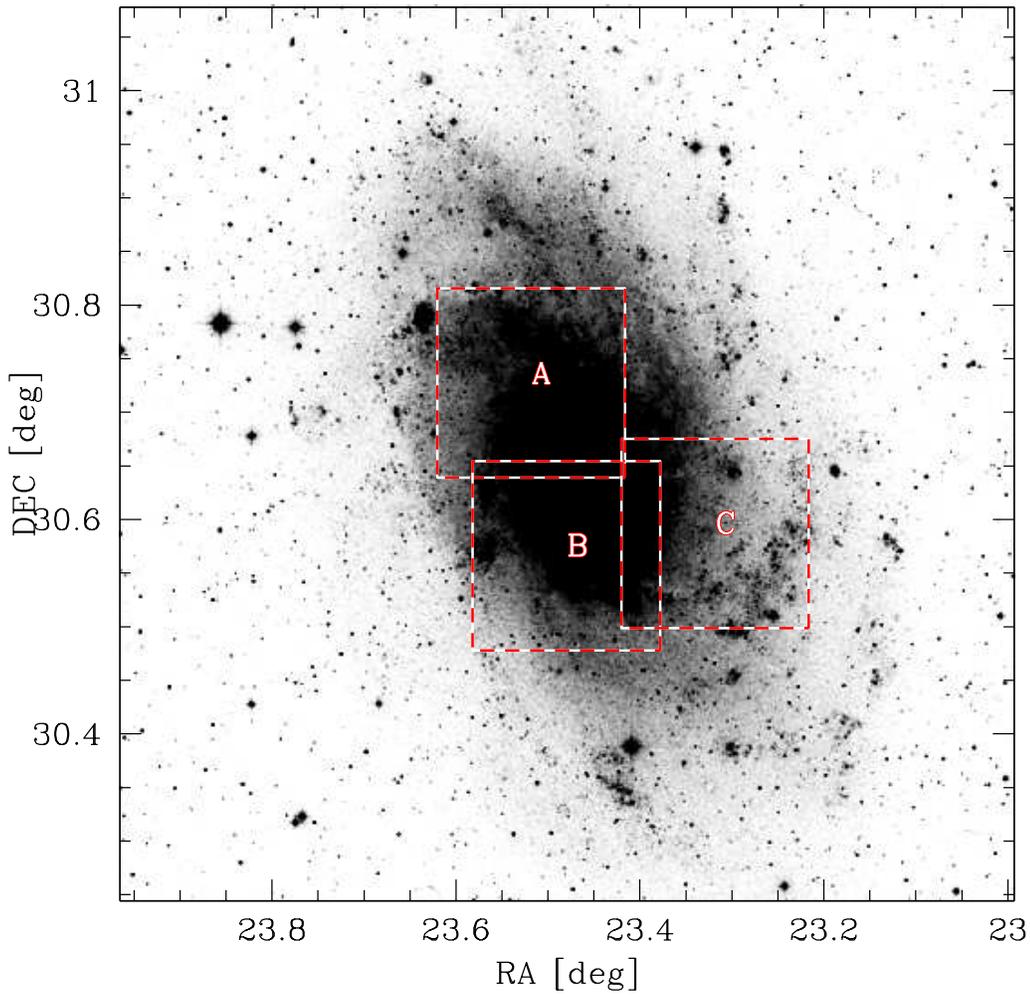} 
\vspace{-3.5cm}
\caption[]{The location of fields A - C in the M33 galaxy}
\label{33f}
\end{figure}

\clearpage

%\input{m31maps.tex}
%\clearpage
%\input{m33maps.tex}
%\clearpage

\end{document}